\documentclass[pre,twocolumn,showpacs]{revtex4}

\usepackage{amsmath}
\usepackage{amssymb}
\usepackage{graphicx}

\begin{document}
\narrowtext
\title{The agents' preferences and the topology of networks}

\author{Daniel O. Cajueiro}
 \affiliation{Department of Economics, Catholic University of Brasilia, 70790-160,
Brasilia, DF, Brazil.}
\begin{abstract}
In this paper, a new framework to study weighed networks is
introduced. The idea behind this methodology is to consider that
each node of the network is an agent that desires to satisfy
his/her preferences in an economic sense. Moreover, the formation
of a link between two agents depends on the benefits and costs
associated to this link. Therefore, an edge between two given
nodes will only arise if the tradeoff between satisfaction and
cost for building it is jointly positive. Using a computational
framework, I intend to show that depending on the agents
combination of benefits and costs, some very well known networks
can naturally arise.

\end{abstract}


 \maketitle

During recent years, one of the main issues of the statistical
physics literature has been the study of dynamic systems such as
airports, wireless links, financial institutions, web pages and
other communication networks and social networks that may be
described by complex weblike structures \footnote{A comprehensive
review of this literature may be found in~\cite{albbar02}}.

On one hand, several models such as small world
networks~\cite{watstr98,wat99} and free scale
networks~\cite{baralb99} have been introduced to specially
accommodate the particularities of these structures that could not
be modeled by the seminal well known random
graphs~\cite{erdren60}. One should notice that although most
attempts have been devoted to the study of unweighed undirected
networks as the ones presented in~\cite{watstr98,baralb99},
recently some researchers have also introduced models to deal with
undirected weighted networks~\cite{yoojeo01} and also directed
digraphs~\cite{krarod01}.

On the other hand, several measures have been presented aiming at
characterizing the properties of these networked systems, for
instance, characteristic path length~\cite{wie47}, clustering
coefficient~\cite{watstr98}, efficiency~\cite{latmar01,latmar02},
cost~\cite{latmar02}, node degree~\cite{baralb99}, degree
correlation~\cite{pasvaz01}, weighted connectivity
strength~\cite{yoojeo01} and disparity~\cite{barbar05}. The main
advantage of using these measures to analyze these complex
structures is the ability to compare different systems with each
other and also to develop a unified theory to approach these
systems.

This paper focuses particularly on undirected weighted graphs. It
proposes another way based on economic and decision theory to cope
with these systems. I suppose that each node of the network is an
agent~\footnote{Throughout this paper nodes and agents are
synonymous.} that has his/her own preferences and is ``starving''
to maximize them. Since all agents in the network will interact in
order to maximize their preferences, an edge between two given
nodes will only arise if the tradeoff between satisfaction and
cost for building it is jointly positive. It is assumed that this
happens when the benefit brought to an agent is greater than his
own cost and the cost left by the other agent (that sometimes is
zero). Therefore, if the benefits brought to the agents by the
edge are positive enough to compensate the cost of construction,
then the edge will exist. This makes sense if one considers that a
connection between agents always brings some kind of benefits, but
the connection sometimes does not exist in a given network because
of the high costs involved.

This tradeoff just presented above is very related to the
formalism developed by~\cite{latmar01,latmar02} since the authors
also seek a tradeoff between satisfaction (measured in a very
specific way as efficiency of communication between the nodes) and
cost (also measured in a very specific way)~\footnote{Actually,
these ideas were borrowed from engineering and operations research
where researchers have been studying optimal paths in networks for
a long time in order to maximize some measure of efficiency and/or
minimize some measure of cost. These attempts were responsible for
the arising of the seminal problems such as the minimum spanning
tree problem, shortest path problem, maximum flow problem etc. A
review of these seminal problems may be found in~\cite{hillie01}.
However, although in~\cite{latmar01,latmar02} there is a similar
flavor, the motivation here is totally different. I am not
directly interested in characterizing the network topology by
measuring its properties and the center of attention here is not
necessarily small world networks. Moreover, the reference of the
``best'' network here is not necessarily the complete network,
because it simply may not be the network that maximizes agent
preferences.}.

Preferences here are modelled as in the economic or decision
theory as utility functions. Specifically, I considere that each
agent has an utility function given by

\begin{equation}u_i(G)=\sum_{\forall j\in \mathcal{N}(G)\backslash i} a_{ij}( w_{ij}-c_{ij})\;\;\;\;\forall i\in G\label{eq:linear}\end{equation}
where $\mathcal{N}(G)$ is the set of nodes in a graph (network)
$G$, $A=[a_{ij}]$ is the adjacency matrix, $W=[w_{ij}]$ is the
matrix of weights and $C=[c_{ij}]$ is the matrix of costs.

In this context, I am particularly interested in the networks that
are the solution to the problem

\begin{equation}\max_A{\sum_{i\in \mathcal{N}(G)}u_i(G) } \label{eq:util}\end{equation}

Therefore, this paper does not approach the mechanisms of networks
formation but it seeks the best topology for a given set of
parameters.

The concept of ``efficiency'' provided by equation
(\ref{eq:util}), which focuses on the total ``productivity'' of
the network and how this allocation is made among individual
agents~\footnote{Considering the simple formulation of equation
(\ref{eq:linear}), this notion is also a Paretian one.}, is the
same one used in~\cite{jacwol96,dutmut97,balgoy00,jac01} to
approach-- in a game theoretical framework-- the dynamics of
network formation and the relation between the concepts of
efficiency (introduced above) and stability~\footnote{The
definition of a stable network comes from the thought that agents
have the discretion to form or reject links. The formation of a
link requires the consent of both parties involved, but severance
can be done unilaterally. This concept is not considered here.}.

The focus of this paper, differently
from~\cite{jacwol96,dutmut97,balgoy00,jac01},  is to provide a
computational framework to relate agent preferences to network
topologies. Thus, one has to maximize equation (\ref{eq:util}) to
reach the desired solution \footnote{This is not the first time
that a kind of maximization principle is used to understand the
topology of complex networks. In \cite{rodrin92}, coping with
natural drainage networks, it is showed that fractal and
multifractal properties evolve from arbitrary initial conditions
by minimizing the local and global rates of energy expenditure in
the system.}. One should notice that since equation
(\ref{eq:util}) has been specified as a linear function, this can
be solved as a linear binary programming problem.

\emph{Binary linear programming} Binary linear programming is a
problem very well studied in the field of operations research and
there are several methods to solve it. Unfortunately, however, due
to its combinatorial nature, this problem is not trivially solved.
Sometimes due to its computational cost, the size of the problem
is constrained or an heuristic method that can provide only a
sub-optimal solution instead of an optimal one is used.

In this paper, since there are no constraints and, in equation
(\ref{eq:linear}), the choice of edges are independent of each
other, the solution of (\ref{eq:util}) is trivial
\footnote{However, in the general case, the branch and bound
technique~\cite{bal65,vanwol87,crojoh83,johkos85} is usually
considered. The basic concept underlying this technique is to
divide and conquer. Since the original ``large'' problem is so
difficult to be solved directly, it is divided into smaller
subproblems until these problems can be conquered  -- this is the
branch step. The conquering step is done partially by bounding how
good the best solution in the subset can be and then discarding
the subset if its bound indicates that the optimal solution is not
in it. A detailed revision of the methods may be found
in~\cite{hillie01}.}.

\begin{figure}
\includegraphics[width=7cm,height=7cm]{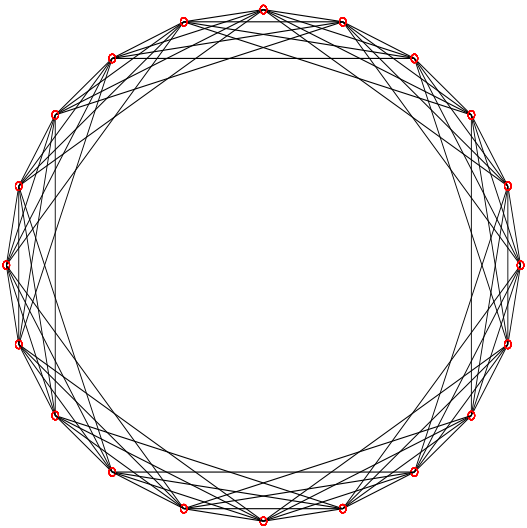}
\caption{A typical regular lattice that arises with $n=20$ and
$K=8$.} \label{fig:redereg}
\end{figure}

\emph{Lattices with $K$ neighbors.} The arising of a regular
network where each node has  $K$ neighbors as a solution of
problem (\ref{eq:util}) is in general only possible if all the
agents have homogeneous preferences with constant benefits over
all agents and a cost that depends only on some measure of the
distance between them (not necessarily physical distance). In
spite of the latter hypothesis being reasonable in the real world,
the former is very hard, since agents in general have different
interests. If the agents are labelled with ordinal indices from
$1$ to $n$, where $n$ is the number of nodes, without loss of
generality, one may suppose in this case that
\begin{equation} w_{ij}=\frac{K}{2\;\mathrm{floor}(n/2)}\end{equation}
and
\begin{equation}c_{ij}=\frac{\min{(|i-j|,n-|i-j|)}}{\mathrm{floor}(n/2)}\end{equation}
where $\mathrm{floor}(x)$ is a function that evaluates the biggest
integer less than $x$ and $|x|$ is the absolute value of $x$. A
typical lattice that arises in this case when $n=20$ and $K=8$ is
shown in figure \ref{fig:redereg}.

\begin{figure}
\includegraphics[width=7cm,height=7cm]{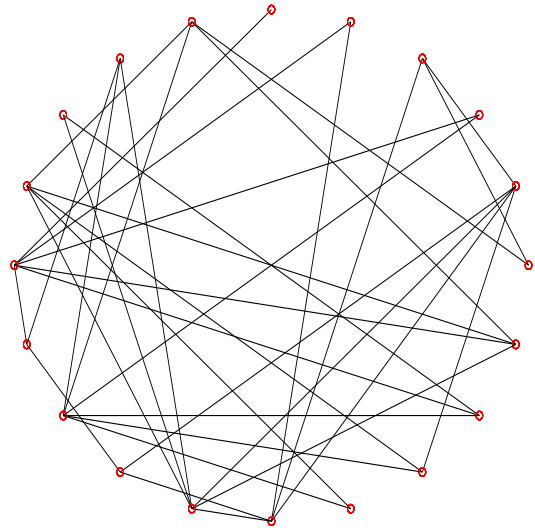}
\caption{A typical random graph that arises with $n=20$ and
$p=0.2$.} \label{fig:rederand}
\end{figure}

\emph{Random Graphs}. Random graphs are the opposite of regular
lattices with $k$ neighbors. The agents take random preferences
into account. This specially works if the benefits brought by the
connections between two nodes are random with magnitude given by a
variable $p$ and the cost of building this connection is constant
as, for instance,

\begin{equation}w_{ij}=p+\epsilon\end{equation}
and
\begin{equation}c_{ij}=1 \end{equation}
where $p$ is the probability of an edge connecting nodes $i,j\in
\mathcal{N}(g)$ and $\epsilon$ is random variable with uniform
distribution in the set $[0,1]$. A typical network that arises in
this case when one solves (\ref{eq:util}) with $n=20$ and $p=0.2$
is shown in figure \ref{fig:rederand}.

Again, as in the case of the regular lattices, this kind of
network is not likely to arise in real life due to the constante
cost.

\emph{Small Worlds.} If one leaves the two extremes presented
above, as in~\cite{watstr98,wat99}, one may arrive at small world
networks. Therefore, one should now consider a set of agents that
with probability $p$ the connection with another agent in the
network brings a benefit modelled by a random variable $\epsilon$
with uniform distribution in the set $[0,1]$ and that with
probability (1-p) the benefit is given by a constant. The first
mechanism described above models the unusual phenomenon of
receiving a large benefit from a distant agent or not receiving a
good benefit from a close agent. The latter mechanism models the
usual phenomenon of receiving a good mechanism from a close agent.
Additionally, as in real life the cost of establishing a
connection depends on some measure of distance.

Mathematically, with probability $p$
\begin{equation}w_{ij}=\epsilon \end{equation}
where $\epsilon$ is a random variable with uniform distribution in
the set $[0,1]$ and with probability $(1-p)$
\begin{equation} w_{ij}=\frac{K}{2\;\mathrm{floor}(n/2)}\label{eq:bensw}\end{equation}
On the other hand,
\begin{equation}c_{ij}=\frac{\min{(|i-j|,n-|i-j|)}}{\mathrm{floor}(n/2)}\label{eq:costsw}\end{equation}
Therefore, the solution of equation (\ref{eq:util}) provides a
network with small world behavior.

\begin{figure}[t]
\includegraphics[width=7cm,height=7cm]{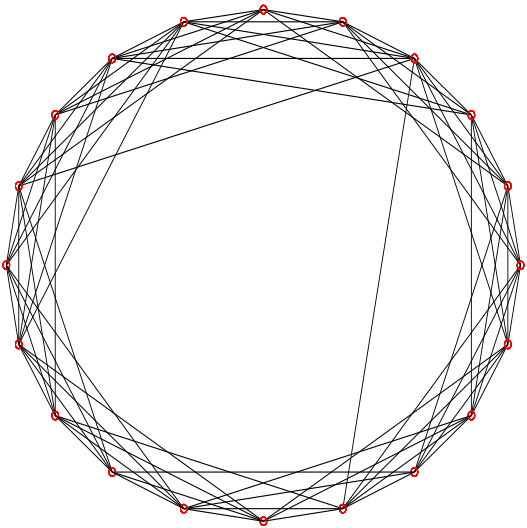}
\caption{A typical small world that arises with $n=20$, $K=8$ and
$p=0.2$.} \label{fig:redesw}
\end{figure}

As we know, several examples of real networks follow this kind of
behavior. If one analyzes the preferences of the agents, it makes
sense. An agent, for example, receives constant benefits (in
average) from being connected to other agents, but there are some
agents who receive lower or bigger benefits than the average. In
figure \ref{fig:redesw}, a typical small world that arises in this
case when one solves (\ref{eq:util}) with $n=20$, $K=8$ and
$p=0.2$ is shown.

\begin{figure}[t]
\includegraphics[width=7cm,height=7cm]{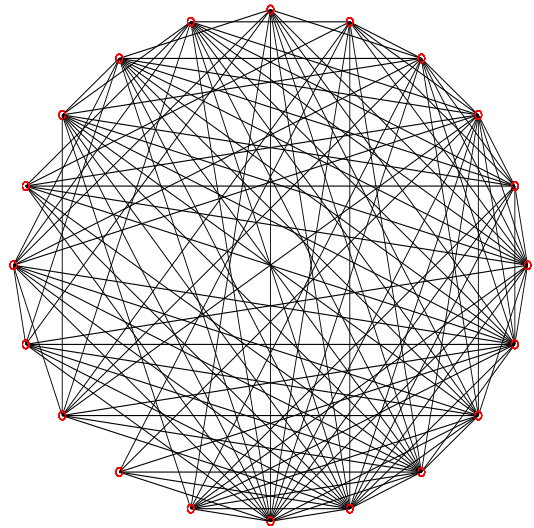}
\caption{A typical free scale network that arises when $n=20$.}
\label{fig:redefs}
\end{figure}

\emph{Free Scale Networks.} Differently from the other situations
considered in this paper, the phenomenon behind the generation of
free scale networks seems to be a kind of cost hierarchy between
the nodes, i.e, there are some nodes that are less costly than the
others. Therefore, some agents will preferentially attach to these
nodes. More specifically, without loss of generality, let $w_{ij}$
and $c_{ij}$ be defined as
\begin{equation}w_{ij}=\epsilon \label{eq:benfree}\end{equation}
$\epsilon$ is a random variable with uniform distribution in the
set $[0,1]$ and
\begin{equation}c_{ij}=\frac{i}{n}\label{eq:costfree} \end{equation}
In equation (\ref{eq:costfree}) it was supposed that the nodes
with minor indices are less costly than the others. Hence, these
nodes will likely present the highest degrees in this case. These
networks, like the small worlds networks, are very likely to be
found in real life. One should think for instance of a network of
airports. There are some airports that due to their geographic
locations are less costly than the others. In figure
\ref{fig:redefs}, there is a typical free-scale network that
arises when one solves (\ref{eq:util}) with $n=20$. In fact, one
may clearly notice the preferencial attachment presented in the
network of this figure.

Moreover, simulations with bigger sets like $n=1000$ yielded
networks with $\gamma= 2.4\pm 0.2$ where $\gamma$ is the exponent
of equation $P(k)\sim k^{-\gamma}$ and $k$ is the degree of a node
in the network.

\emph{Final Remarks. }In this note, I have presented a new
computational framework to characterize complex networks, i.e.,
one that may characterize the networks by the preferences of their
agents (nodes). Actually, although only the four most common
classes of networks have been considered, this framework can be
used for many classes. In particular, by mixing the preferences of
the agents presented in equations (\ref{eq:bensw}),
(\ref{eq:costsw}) (\ref{eq:benfree}) and (\ref{eq:costfree}), one
may find networks with small world behavior and also attach
preferences. Moreover, this methodology also works for weighted
digraphs.

On one hand, linear utility functions, which means that the agents
are indifferent to the risk, were the only class of utility
functions considered here. A question that arises is: What effect
is expected in the topology of the networks if the agents are, for
instance, averse to the risk with concave utility
functions~\footnote{Clearly, if the utility functions of the
agents are not linear, linear binary programming cannot be used to
find the optimal solution of this new referred problem, but
another method may be applied. One of the most common choices in
the general situation is the algorithm genetic~\cite{gol89}.}.
Furthermore, no constraint has been considered in the optimization
problem provided by (\ref{eq:util}). What kind of constraints are
the agents in the real world subjected to and what kind of effect
will these constraints cause in the topology of networks?

On the other hand, the matrices $W$ and $C$ here were considered
exogenous, i.e., they were formed prior to the solution of the
problem. It is also possible to suppose that these matrices have
elements that depend on the parameters of a given iteration of the
problem. For instance, the benefit brought by node $i$ to node $j$
could depend on the number of nodes that $i$ actually
possesses~\footnote{Again, this cannot be solved by binary linear
programming, but another method could be applied using the
framework of this paper.}. This could be the root for the study of
network formation using this kind of framework.

In summary, this proposed framework may be used to improve the
understanding of these complex networks that are present
everywhere.

\bibliography{cajueiro_util_rede}

\end{document}